\documentstyle[preprint,prl,aps]{revtex}
\begin{document}
\draft
%
%
\title{Optimal Measurements of Magnetic Flux in Superconducting Circuits\\
and Macroscopic Quantum Mechanics}
\author{Tommaso Calarco${}^{1,2}$ and Roberto Onofrio${}^{1,3}$}
\address{${}^{1}$Dipartimento di Fisica ``G. Galilei'',
Universit\`a di Padova, Via Marzolo 8, Padova, Italy 35131\\
${}^{2}$INFN, Laboratori Nazionali di Legnaro, Legnaro, Italy 35020\\
${}^{3}$INFN, Sezione di Padova, Italy 35131
}
\date{\today}
\maketitle
%
%
\begin{abstract}
A model of repeated quantum measurements of magnetic flux in
superconducting circuits manifesting tunneling is discussed.
The perturbation due to the previous measurements of magnetic
flux is always present unless quantum nondemolition measurements are
performed. By replacing the classical notion of noninvasivity with this
condition, temporal Bell-like inequalities allows one to test the
observability at the macroscopic level of the conflict between
realism and quantum theory.
\end{abstract}
%
%
%
%
\narrowtext
Quantum mechanics, when compared to classical
mechanics, shows two peculiar differences.
Firstly, a coherent structure for the pure states
and their dynamical evolution is present.
Secondly, the evolution of the state of the measured system is
unavoidably affected by the measurement process.
A global understanding of the theory and its relationship to the
classical limit requires to extend its validity to the macroscopic world.
In this domain the conflict between its structure and the classical
sense of physical reality cannot be overcome \cite{BELL} and experiments
aimed to compare the predictions of the two world views are crucial.
With this spirit Leggett and Garg proposed experiments which could help to
discriminate between the two structures in the macroscopic domain \cite{LEGG1}.
They introduced a class of realistic theories which
do not satisfy the two abovementioned peculiarities of quantum mechanics,
both in the sense that ``a macroscopic system with two or more
macroscopically distinct states available to it will at all times {\it be}
in one or the other of these states'' ({\it macroscopic realism}) as well as
that ``it is possible, in principle, to determine the state of the system with
arbitrarily small perturbation on its subsequent dynamics'' ({\it noninvasive
measurability at the macroscopic level}) \cite{LEGG1}.
This leads the authors to a quantitative test introducing
inequalities between correlation functions of observables of a macroscopic
system, a superconducting quantum interferometer device
(SQUID) subjected to a sequence of repeated measurements of magnetic flux.
Besides extending quantum mechanics in a territory dominated so far
by classical mechanics, any possible failure of quantum theory to
describe the results of an experiment of such a kind could give
limits on its validity whenever the intrinsic
complexity of a macroscopic system can play an unknown role \cite{LEGG2,LEGG3}.
Measurement schemes following that path
have been proposed both using a set of SQUIDs \cite{TE}
and four-level atoms and mesoscopic structures \cite{PAZ}.
In \cite{ELLIS} experiments of this kind have been also discussed for
observing topological effects due to the quantized structure of the vacuum.
However, the proposal of Leggett and Garg has been
criticized due both to the interaction with
the external environment \cite{KRAICH} and the limitations given by quantum
mechanics to the repeated measurement of observables \cite{BA1,PE}.
Furthermore, the statement that noninvasivity is simply a consequence of
macrorealism, at the basis of the original argument
(see the reply in \cite{BA1}), has been repeatedly discussed
\cite{GRUPPONE}.
This last obstacle can be overcome by
weakening the classical noninvasivity constraint in the so-called quantum
nondemolition requirement \cite{CAVES,BRAG}, {\em i.e.} by using
{\em quantum} measurement strategies which minimize
the influence of the previous measurements on the
predictability of the next outcomes.
The whole argument remains now valid without any classical
({\em i.e.} incompatible with the Heisenberg uncertainty principle) assumption
on the measurement process, but holds only for particular instants of time.

In this Letter we study quantitatively optimal strategies for repeated
measurements of magnetic flux in bistable potentials which can schematize
the rf-SQUID behaviour.
Such a potential can be written as
\begin{equation}
V(\varphi)=-{\mu\over 2} \varphi^2+{\lambda \over 4} \varphi^4
\label{SOMBRER}
\end{equation}
where $\varphi$ is the trapped magnetic flux,
$\mu$ and $\lambda$ are parameters associated to the superconducting circuit.
Implicit in this assumption is that the rf-SQUID has been
decoupled in such a way to neglect the influence of the external environment.
This allows us to describe the system in terms
of pure states $\psi(\varphi)$, although it has been shown that quantum
mechanical predictions are distinguishable from classical models
also in the more realistic situation of an overdamped regime \cite{CHALEG}.
Since general arguments exist
on the fundamental noise introduced by any linear amplifier \cite{HEFF,HAUS},
the problem of the measurement of flux in a superconducting circuit
is independent upon the detailed scheme used to detect
the quantum state of the rf-SQUID.
This happens also for the so-called null-result measurements, in which
interaction with the instrument must occur leading to quantum noise
fed into the observed system, as outlined in \cite{PE}.

The model we use is based upon the path integral approach and the
restriction imposed by a measurement in the space of the paths \cite{ME}.
The standard quantum limit in a continuous measurement
of position for nonlinear systems has been already analyzed
in the framework of the path-integral approach \cite{MEOP1}.
This approach has been also applied to describe impulsive
measurements in nonlinear systems \cite{MEOP2} and to understand
quantum Zeno effect in atomic spectroscopy \cite{OPTA}.
The measuring system is schematized by an arbitrary
measurement output $\Phi(t)$ and an instrumental error $\Delta \Phi$,
two parameters hopefully present for every meter.
The effect of the measurement modifies the  path-integral giving
privilege to the paths close to the output $\Phi(t)$.
The propagator of a system in which the magnetic flux is measured includes
the influence of the measurement through a weight functional
$w_{[\Phi]}[\varphi]$
\begin{equation}
K_{[\Phi]}(\varphi^{\prime\prime},\tau;\varphi^{\prime},0)=
\int d[\varphi]\exp \biggl\{ {i \over {\hbar}}\int_0^{\tau}L(\varphi,
\dot{\varphi},t)dt
\biggr\}w_{[\Phi]}[\varphi].
\label{PROP}
\end{equation}
If the system is initially in a pure state described by the wavefunction
$\psi(\varphi,0)$ the quantity
\begin{equation}
P_{[\Phi]}={{\vert\langle\psi_{[\Phi]}(\tau)|\psi_{[\Phi]}
(\tau)\rangle\vert}^2
\over {\int {\vert\langle\psi_{[\Phi]}(\tau)|\psi_{[\Phi]}(\tau)\rangle
\vert}^2 d[\Phi]}}
\label{P}
\end{equation}
where
\begin{equation}
\psi_{[\Phi]}(\varphi^{\prime\prime},\tau)={\int}
K_{[\Phi]}(\varphi^{\prime\prime},\tau;
\varphi^{\prime},0)\psi(\varphi^{\prime},0)d\varphi^{\prime}
\label{PSITAU}
\end{equation}
can be interpreted as a probability functional for the measurement output.
Due to the influence of the measurement an effective magnetic flux
uncertainty arises
\begin{equation}
\Delta \Phi_{\it eff}^2=
{2 \int {1 \over \tau} \int_0^{\tau} [\Phi(t)-\tilde{\Phi}(t)]^2 dt~
P_{[\Phi]}d[\Phi] \over
 \int P_{[\Phi]}d[\Phi] }.
\label{DAEFFTH}
\end{equation}
\noindent
where $\tilde{\Phi}(t)$ is the most probable measure path which
makes $P_{[\Phi]}$ extremal. In other words $\Delta \Phi_{\it eff}$
is the minimum dispersion of the path obtained for the most
probable result of the measurement.
In general the effective uncertainty $\Delta \Phi_{\it eff}$
is greater than the instrumental error $\Delta \Phi$, expressing
the spreading of the paths due to the effect of the back-action
of the meter on the measured system,
unless the system is monitored in a regime unaffected by the quantum
noise, {\em i.e.} when $\Delta \Phi \gg \sigma$
where $\sigma$ is the width of
the initial wavefunction $\psi(\varphi,0)$,  or in a QND way
\cite{BRAG,MEOP2}.
For simplicity we represent an actual measurement with instrumental
error $\Delta \Phi$ lasting a time $\tau$ through a weight
functional $w_{[\Phi]}[\varphi]$
\begin{equation}
w_{[\Phi]}[\varphi]=\exp \biggl \{-{1\over 2\Delta \Phi^2 \tau}
\int_0^{\tau} [\varphi(t)-\Phi(t)]^2 dt \biggl \}
\label{WA}
\end{equation}

As shown in \cite{MEOP1},  the evaluation of the path-integral
can be overcome by writing an effective Schr\"odinger
equation which takes into account the influence of the measurement.
This allows to study a strategy which consists of a sequence of measurements of
magnetic flux of duration $\tau$ equally spaced by a quiescent time $\Delta T$
in which no measurements are performed.
The results were obtained in an impulsive regime as defined in \cite{MEOP2},
in such a way that the computed quantities
do not depend upon the duration of the measurement.
In this limit the collapse induced by the propagator (2) can be represented
through the operator
\begin{equation}
\hat{w}_\Phi\equiv \exp\left\{-\frac{(\hat{\varphi}-\Phi)^2}{2\Delta
\Phi^2}\right\}.
\end{equation}
The initial wavefunction
$\psi(\varphi,t_0)= \sum_{j=1}^\infty c_j \psi_j(\varphi)$, where the
$\psi_j(\varphi)$ are the energy eigenstates having eigenvalues $E_j$,
after a sequence of measurements
$\{\Phi_n\}_{n=0,\ldots,N}$ at times $t_n\equiv n \Delta T$ becomes
\begin{equation}
\label{psimis}
\psi_{\{\Phi_n\}}(\varphi,t_N^+)=
\hat{w}_{\Phi_N}
\left(\prod_{j=1}^N
e^{-\frac{i}{\hbar}\hat{H}\Delta T}\hat{w}_{\Phi_{N-j}}
\right)\psi(\varphi,t_0^-).
\end{equation}
With a particular sequence of identical outcomes, $\Phi_n\equiv\Phi_0$ for
$n\leq N-1$, the wavefunction, after insertion of $N$ completeness
relationships, is written as
\begin{equation}
\psi_{\{\Phi_n\}}(\varphi,t_N^+) = \sum_{m,l=1}^\infty
B_{ml}(\Phi_N,\Phi_0,\Delta T,N,\Delta \Phi) c_l \psi_m(\varphi),
\end{equation}
where
\begin{equation}
B_{ml}= \sum_{\left\{n_j\right\}}^\infty
W_{mn_1}^{\Phi_N}
\left(\prod_{j=1}^{N-1}W_{n_jn_{j+1}}^{\Phi_0}\right)
W_{n_Nl}^{\Phi_0} e^{-\frac{i}{\hbar}\Delta T
\sum_{j=1}^{N}E_{n_j}},
\end{equation}
and the $W_{ij}^\Phi$ are the matrix elements of $\hat{w}_\Phi$ on the
$\psi_j$'s.
This allows to calculate, through (\ref{P}), the effective magnetic flux
uncertainty (\ref{DAEFFTH}).
We have checked that this technique gives
the same results of the previous one described in \cite{MEOP2}
and allows to reduce the computation time by several orders of magnitude
obtaining the same accuracy.
Initially we have chosen a square-well of width $2\bar{\Phi}$ with a
delta-like potential at the center, having eigenstates and eigenvalues
analytically calculable (see for instance \cite{SEGRE}).
Besides the possibility to control the dynamical evolution
in an analytic way the presence of a unique parameter, the height of the
central barrier $V_0$, allows us  to simply discuss in detail a number of
features common to all the topologically equivalent
potentials - like the one described by (1) - which exhibit bistability.
 As already discussed in \cite{MEOP2} the optimality of the measurement
is dictated by the spectral properties of the system.
The asymptotic collapsed wavefunction can be expanded in terms
of the eigenstates of the potential and the optimal measurement
is obtained provided that the quiescent time is commensurable
to the characteristic times for the wavefunction reformation
\begin{equation}
\label{tij}
T_{ij}\equiv\frac{2\pi\hbar}{|E_i-E_j|},
\end{equation}
where $E_i$, $E_j$ are the energy eigenvalues which have maximal
projections on the asymptotic state.
The two eigenstates which maximally contribute to the wavefunction
reformation after tunneling are the two states corresponding to
the first and the second eigenvalue, whose splitting is related to
the tunneling period.

In Fig. 1 we show the dependence of the effective uncertainty of the magnetic
flux $\Delta \Phi_{\it eff}$ upon the number of measurements for different
values of the quiescent time $\Delta T$.
It is evident that $\Delta \Phi_{\it eff}$ reaches an asymptotic value after
few collapses around the measurement result, chosen
to be $\Phi_0\equiv\bar{\Phi}/2$, which has been verified to be
the most probable result.
The asymptotic $\Delta \Phi_{\it eff}$ depends upon
the quiescent time $\Delta T$.
As we show in Fig. 2 the asymptotic $\Delta \Phi_{\it eff}$ has minima
close to $\Delta \Phi$ when $\Delta T$ is a multiple
of the characteristic period $T_{12}$.
Unless the measurements are repeated with this periodicity noise due to the
measurement process is fed into the system
affecting the following measurements.
In the limit of an impenetrable barrier the energy levels get twice degenerate
and the even (odd) eigenstates are simply (anti)symmetric
combinations of the half-well ones.
Thus the spectrum of a wave packet symmetrically
centered in the middle of one half-well shall not contain
the eigenvalues $E_{4n}, E_{4n-1}$.
For $V_0$ finite but high, $T_{26}$ -~or equivalently $T_{15}$~- is
therefore expected to be a good choice for the optimal quiescent time,
as shown in Fig. 3, a magnification of Fig. 2 for short times.
The periodicity linked to the intrawell time $T_{26}$ is evident.
This optimal period becomes increasingly important in the limit of opaque
barriers, as shown in the insert of Fig. 3, where the optimal
$\Delta \Phi_{\it eff}$ is plotted  versus
the height of the potential barrier $V_0$.
Other cases with analytically calculable eigenstates and eigenvalues,
namely the double harmonic oscillator and the square well with a
rectangular barrier in the middle, have been also studied,
as we will report in detail in a forthcoming paper.
In Fig. 4 we show the results obtained for the more realistic case of
the potential of Eq. (1).
In this case we have numerically evaluated the eigenstates
using a relaxation algorithm based upon repeated collapses of a trial
wavefunction subjected to measurements of energy \cite{OPTA,PRETA}
around the eigenvalues estimated with the WKB method.
The behaviour does not differ from the case shown in Fig. 2,
and the optimal period is still the tunneling time.
The method allows us to describe a general set of repeated measurements.
We have calculated what happens in the case of three consecutive
measurements of magnetic flux, a case of crucial interest for
the experiments on temporal Bell inequalities.
An optimal set of measurements prepares the state allowing to define
an asymptotic effective uncertainty approximately equal to the classical
one, and after that the
same quantity is calculated for a specific sequence of measurements.
In the particular example in Fig. 5a) we show the result of this analysis
for the case of two measurements, both with negative result,
performed at times $t_a$ and $t_c$ (no measurement being performed at the
intermediate time $t_b$). This corresponds to study the quantum limit in the
sequence of measurements whose correlation probability is
\begin{equation}
P^{ac}_{--}=P\{\Phi(t_a) < 0, \Phi(t_c) < 0\},
\label{Q}
\end{equation}
which has relevance for the study of quantum-mechanically predicted
violations to the temporal Bell-like inequalities, as in \cite{LEGG1}.
It is evident in our particular example that the minimum effective
uncertainty is obtained for quiescent times such that $\Delta t_{ab}+
\Delta t_{bc}$
($\Delta t_{ab}=t_b-t_a$, $\Delta t_{bc}=t_c-t_b$)
is a multiple of the optimal periodicity $T_{12}$ already individuated
for a sequence of repeated measurements with the same result.
We show also, in Fig. 5b), the behaviour of the effective uncertainty
corresponding to $P^{bc}_{+-}$, defined as above.
In all the considered cases it is evident that
the predictability of the measurement outcome
is affected by the uncertainty due to the previous measurements.
One can minimize this by simply using optimal quiescent times which are
analytically determined through (\ref{tij})
once the eigenvalues and the eigenstates are given.
The particular Bell-type inequalities which involve correlation functions
between measurements performed at {\em these} times
hold {\em without} any demand such as noninvasivity: therefore if the
Bell inequalities were not experimentally violated in {\em these} cases, this
would not be attributable to the Heisenberg uncertainty principle and
could represent a {\em direct} confirmation of the realistic hypothesis.

Correlations between measurements of magnetic flux appearing in
temporal Bell inequalities can be calculated in our approach
according to  (\ref{P},\ref{Q}).
Thus we can verify for whose quiescent times quantum mechanics,
including the measurement process, predicts violations of
Bell inequalities. Observation of these last will be meaningful only if
the quantum uncertainty resulting from the sequence of measurements still
allows one to resolve the
two half-wells, {\em i.e.} to identify two macroscopically distinguishable
states. This will be the subject of a future paper \cite{BELLFUT}.

\acknowledgments
We acknowledge C. Presilla for crucial discussions and G. C. Ghirardi
for a critical reading of the manuscript.
One of us (T.C.) is grateful to M. Cerdonio and S. Vitale
for the kind hospitality at the University of Trento.
%
%

%
%
\begin{figure}
\caption{Effective uncertainty $\Delta \Phi_{\it eff}$ versus the number
$n$ of repeated impulsive measurements of magnetic flux
in a square-well with a $\delta$-like potential at the center
for different quiescent times, expressed in terms of the periodicity $T_{12}$.
The result of the measurement is always taken as $\Phi_n=\bar{\Phi}/2$.
We put $2m=\hbar=\omega=1$, $\Delta \Phi=1$.}
\end{figure}

\begin{figure}
\caption{Dependence of the asymptotic effective
uncertainty $\Delta \Phi_{\it eff}$
on the quiescent time $\Delta T$ for the $\delta$-like potential
in the square well.
We put $V_0=500, \Delta \Phi=2$.}
\end{figure}

\begin{figure}
\caption{Dependence of the asymptotic effective uncertainty
$\Delta \Phi_{\it eff}$
on the quiescent time $\Delta T$ on a magnified scale comparable to the
characteristic time $T_{26} \ll T_{12}$. In the insert the dependence upon
$V_0$ is depicted for $\Delta T=T_{26}$, showing that this
periodicity is optimal in the limit of decoupled half wells.}
\end{figure}

\begin{figure}
\caption{Dependence of the asymptotic effective uncertainty $\Delta \Phi_{\it
eff}$ on the quiescent time $\Delta T$ for the potential in (1). In this
example we have $\mu=9.6$ and $\lambda=4.382$ in such a way that 6 energy
levels lie below the central barrier.}
\end{figure}

\begin{figure}
\caption{Dependence of the effective uncertainties $\Delta\Phi_{--}^{ac}$
(a) and $\Delta\Phi_{+-}^{bc}$ (b) on the quiescent times $\Delta t_{ab}$
and $\Delta t_{bc}$
in the case of two sequences, each of three measurements,
and a potential as in (1).
In sequence a), the magnetic flux is measured at time $t_a$ and
$t_c$ and not at $t_b$, and both the measurements are assumed to give
as result a negative magnetic flux;
in sequence b), the magnetic flux is measured at time $t_b$ and
$t_c$ and not at $t_a$, and the measurements are assumed to give
as results respectively a positive and a negative value.}
\end{figure}

\end{document}